\documentclass[twocolumn,showpacs,preprintnumbers,amsmath,amssymb]{revtex4}
\usepackage{graphicx}% Include figure files
\usepackage{dcolumn}% Align table columns on decimal point
\usepackage{bm}% bold math

\begin{document}

\title{Rejuvenation and overaging in a colloidal glass under shear}
\author{Virgile Viasnoff}
\affiliation{
L.P.M UMR7615 CNRS ESPCI 10 rue Vauquelin 75231 Paris, France
}
\author{Fran\c{c}ois Lequeux}
\affiliation{
L.P.M UMR7615 CNRS ESPCI 10 rue Vauquelin 75231 Paris, France
}

\begin{abstract}
We report the modifications of the microscopic dynamics of a colloidal glass
submitted to shear. We use multispeckle diffusing wave spectroscopy to
monitor the evolution of the spontaneous slow relaxation processes after the
sample have been submitted to various straining. We show that high
shear rejuvenates the system and accelerates its dynamics whereas moderate
shear overage the system. We analyze this phenomena within the frame of the Bouchaud's
trap model.
\end{abstract}
\date{\today}
\maketitle

The physical properties of glassy systems such as supercooled liquids, spin
glasses, amorphous polymers and colloidal glasses are well known to evolve
slowly with time. This phenomena is called aging. The out-of-equilibrium
nature of theses systems compels their physical properties to depend on two
times as shown both by theoretical and experimental studies. The first time
is the age of the system i.e. the time spent in the glassy phase. The second
time is the time elapsed since the measurement started. Consequently, a well
controlled history is a key requirement to obtain reproducible results. The
most common way to control the history as the age of the system is to quench
it from an equilibrium state at high temperature into an aging state at low
temperature. Since the system is at equilibrium at high temperature, all the
history preceding the quench is erased and a complete rejuvenation of the
physical properties is achieved. For colloidal glasses however, temperature
may not be a practical parameter. Liu and Nagel recently suggested \cite
{LiuNagel} that shear may act equivalently to temperature for such
materials. Indeed, a high shear proves to be able to erased the memory for
theses systems and thus completely rejuvenate them \cite{cloitre1}\cite
{CaroFranc}. In that sense the cessation of a shear is similar to a
temperature quench. Moreover, different approaches were recently introduced
to describe the coupling between mechanical deformations and aging
phenomenon \cite{solirevue} \cite{Berthier} . However quantitative
experiments are still lacking to determine unambiguously how shear acts on a
microscopic level and how it should be introduced in a mean field model.

In this Letter we report non trivial shear effects on a dense solution of
polybeads. We show that theses effects can be mimicked by temperature
changes in the Bouchaud's trap model \cite{monthus}. Our underlying physical
picture is the following: slow relaxations, of characteristic time $\tau $,
are determined by the structural rearrangements of the particles. The
dynamics slows down with the age $t_{w}$ of the system as the beads find
more and more stable configurations. $\tau $ is thus an increasing function
of $t_{w}$. Since a shear flow seems to be able to rejuvenate completely the
system, one could imagine that it shuffles the beads arrangements. The
resulting configurations could be less stable. The dynamics of rearrangements
would be then accelerated and the relaxation time $\tau $ would decrease.
Oppositely, one could imagine that a moderate oscillatory strain is able to
help the system to find more stable, though always non-crystalline,
configurations. The dynamics would be then slowed down and $\tau $ would be
increased. In order to elucidate theses two contradictory pictures, we
experimentally tested the effect of an oscillating shear strain on the
evolution of the microscopic dynamics of our dense suspension. The sample is
a commercial suspension of polystyrene spherical beads of diameter 162 nm
copoplymerized with acrilic acid (1\%) that creates a charged corona
stabilizing the microspheres. It is concentrated by dialyzisis to a volume
fraction $\varphi =49\%$. We use Multispeckle Diffusing Wave Spectroscopy
(MSDWS) to probe the slow relaxation dynamics of the system after various
strain histories. MSDWS is an extension of regular DWS, a technique that
measures the average displacement of the particles through the intensity
fluctuations of multiply scattered light. Whereas DWS performs a time
average of the fluctuations, MSDWS makes a spatial average of them. It is
thus a well suited technique to study slow transient phenomena such as aging
processes. A precise description of the technique can be found in \cite
{Experimental}. It allows to measure in real time the two times intensity
autocorrelation function $g_{2}(t_{w}+t,t_{w})=$ $\frac{\langle
I(t_{w}+t)I(t_{w})\rangle }{\langle I(t_{w})\rangle ^{2}}$ where $t_{w}$ is
the reference time and t the elapsed time since $t_{w}.$ The average $%
\langle ...\rangle $ is spatially performed over the speckle pattern. This
correlation function is a decreasing function of the number of
rearrangements that occurred between $t_{w}$ and $t_{w}+t$ as demonstrated
in \cite{DurianPine}\cite{CohenAddad}. Thus, the principle of the whole
experiment is the following: the suspension is submitted to different shear
history detailed below; the modification of its dynamical properties is
recorded after shear cessation. The sample is placed in a custom-made shear
cell consisting in two parallel glass plates with a variable gap. For all
presented experiments, the gap was set to 1.3 mm. Oscillatory straining was
realized by moving the bottom plate thanks to a piezoelectric device. Shear
strain from 30\% to 0.04\% could be possibly applied at a fixed frequency of
1Hz. The shear cell was synchronized with the light scattering detection via
a PC. For optical considerations, backscattering geometry was used. We
confirmed that the suspension did not slip on the wall by checking that we
obtained identical results for various gap size. Moreover, no macroscopic
crystallization was observed. All the experiments were performed at room
temperature.  In order to increase the signal to noise ratio, each test
presented in this paper was performed 10 times and each correlation function
was averaged over the 10 experiments. The reproducibility was check to be
better than 5\%.

We first submit the sample to a series of 40 oscillations for different
strain amplitudes $\gamma .$ For $\gamma >20\%$ the measurement after the
shear cessation becomes insensitive to the shear strain amplitude showing
that the rejuvenation is then total. The age $t_{w}$ of the system is
defined from the shear cessation as usually done for temperature quench. Fig
1a shows the correlation function versus t for different values of $t_{w}$%
.This set of curves displays two important features: on the one hand, in the
region where t$\ll 5.10^{-2}s$, all the curves overlap. The correlation
functions show an initial decrease that is the end of a short time
relaxation. It corresponds to restricted thermal fluctuations of the
particles and is called $\beta $ mode. This fast mode is not affected by
shear as predicted by models \cite{Berthier}. On the other hand, in the long
time limit, we observe a slow decay, known as $\alpha $ relaxation, typical
of glasses. This decay from the pseudo plateau region is all the slower as
the system is elder. It thus means that the average rate of the structural
rearrangements decreases with $t_{w}.$ We arbitrarily define the structural
relaxation time $\tau _{1/2}$for this regime so that $g_{2}(t_{w}+\tau
_{1/2},t_{w})-1=0.06$. Fig 1b shows that $\tau _{1/2}\varpropto
t_{w}^{1.06\pm 0.08}$ for $t_{w}>1s.$ The aging part of the correlation
function can be rescaled with the reduced variable $\frac{1}{1-\mu }\left(
(t+t_{w})^{1-\mu }-t_{w}^{1-\mu }\right) $ with $\mu =1.06.$ In addition,
the shape of the correlation function is invariant by such a scaling. This
result is a typical feature of aging process and is qualitatively similar to
that found for rheology of such systems \cite{cloitre1}\cite{CaroFranc}\cite
{Cipelonion}.

\begin{figure}[th]
\begin{center}
\includegraphics[width=8cm,clip]{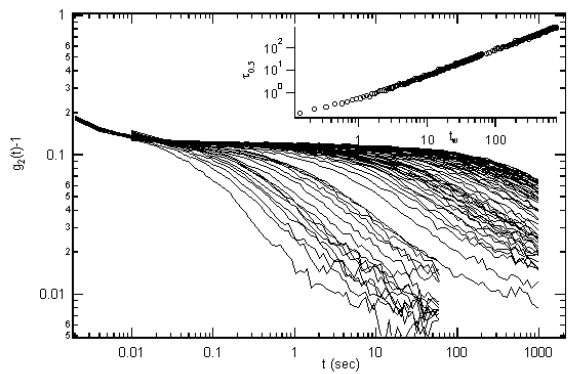}
\end{center}
\caption{The intensity autocorrelation function $g_{2}(t_{w}+t,t_{w})-1$
for different t$_{w}$ ranging from 0.5 s to 10$^{3}$ s. The first decrease
at short times comes from the tail of the $\beta $ relaxation. The long term
decrease is due to the structural relaxation. The insert shows $\tau _{1/2}$
in seconds vs t$_{w}.$ $\tau _{1/2}$ scales as t$_{w}^{1.06\pm 0.08}$ }
\label{fig:courberef}
\end{figure}

In order to better understand the influence of shear on the dynamics of the
particles we apply to the system the strain history described in fig 2a. The
sample is first submitted to an oscillatory strain of amplitude 30\% at 1Hz
during 40 s in order to rejuvenate it totally. Secondly, we let it age at
rest for 10s. Then, a second burst of 1Hz oscillations is applied during t$%
_{d}$. After its cessation, we examine how the amplitude $\gamma $ and the
duration t$_{d}$ of the burst have modified the dynamical properties of the
sample. t$_{w}$ is now referenced from the end of the second burst.

Fig 2b -resp fig 2c- displays the relaxation time $\tau _{1/2\text{ }}$ as a
function of $t_{w}$ for different strain amplitudes, with t$_{d}$=1s -resp  t%
$_{d}$=100s. Two limit cases can be considered: that of a complete
rejuvenation during the second burst -corresponding to the reference curve
of fig1- and that of $\gamma =0\%$ where the system is unperturbed during
the second burst. That last curve is the same than the completely
rejuvenated one but shifted in time by t$_{s}=$10s+t$_{d}$. These two limit
curves merge at long time because of the log scale. For a duration of the
second burst t$_{d}$=1s -fig 2b- we observe that the effect of the second
burst is to rejuvenate \textit{partially} the system: for any $t_{w}$ the
relaxation time is a monotonically decreasing function of $\gamma $. All the
curves lie between the two limit case curves. This is coherent with the idea
that the shear has to be strong enough to rejuvenate totally the system.
However, one might expect that the amplitude is not the only relevant
parameter, but the duration of the shear application has to play a role.
Indeed, if the system is left longer under shear, the modification of its
dynamic is then dual. When t$_{d}=100s$ -fig 2c-, in the limit of short $%
t_{w}$, the relaxation time behaves similarly as previously described.
However, for longer $t_{w}$ the relaxation time after a moderate strain (see
e.g. $\gamma =7\%$) is surprisingly longer than the one for the sample
without solicitation during the second burst. In other words moderate shear
strain results in a system with a slower relaxation time. We call this
overshoot in the relaxation time \textit{overaging}. For large strain
amplitudes a total rejuvenation is recovered as exemplified by the curve for 
$\gamma =11.7\%$ in fig 2c. This experiment demonstrates that a moderate
shear can perturb the dynamical properties of the system in a non-trivial
way.

\begin{figure}[ht]
\begin{center}
\includegraphics[width=8cm,clip]{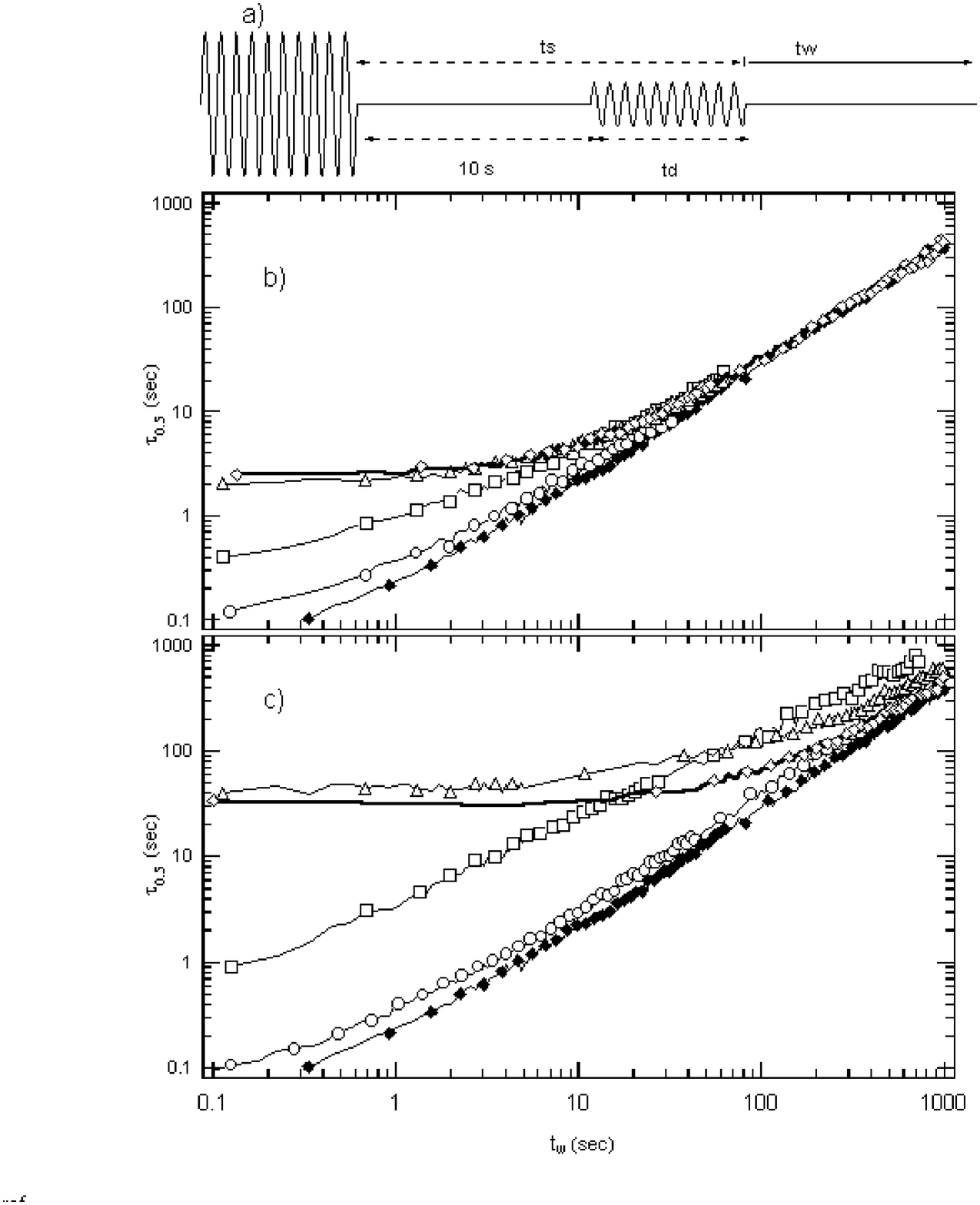}
\end{center}
\caption{ a) Strain history. b) $\tau _{1/2}$ for t$_{s}=11s$ and different
strain amplitudes $\gamma =$ 0\% (Diamond  ), 2.9\% ($\triangle $), 7.9\%
($\square $), 11.7\% (O) and the reference curve ($\blacklozenge $ )$.$ $%
\tau _{1/2}$ decreases monotonically with the strain amplitude at short
times. for t$_{w}\sim t_{s}$ all curve merges. c) $\tau _{1/2}$ for t$%
_{s}=110s$ for the same strain amplitude. $\tau _{1/2}$ for $\gamma =$ 2.9\%
and 7.9\% is superior to $\gamma =$ 0\% in the long time regime. }
\label{fig:sinusoidal}
\end{figure}

This is contradictory with the simple idea that strain or stress simply
rejuvenates the system and accelerates the dynamics. It shows that a
transient strain changes not only the average value of the relaxation time
but also the distribution of relaxation times within the sample. The change
of the distribution is clearly demonstrated by the crossing of the curves in
fig 2c: two similar systems with the same relaxation time and different
histories can evolve differently. The shape of the correlation function is
also altered by the change in the distribution of relaxation time as
demonstrated in fig 4a. Similar modifications of the shape of the response
function have already been noticed on spin glasses after a temperature step
- see fig 5 of ref \cite{Revueverrespin} . We remark that microscopic aging
models including shear in their equations tackles the problem of the
modification of the dynamics under shear. To our knowledge, no calculation
of the modification of the microscopic dynamic after a transient shear has
been performed for theses models. This problem will be addressed in future
work \cite{discussiondetaillee}.

However, we point out here, the similarity of overaging after a transient
shear stress and that predicted by the trap model \cite{monthus} after a
temperature step. We thus introduce in this model the oscillatory straining
as a change of temperature. This model describes the motion of non
interacting particles hopping in an energy landscape with wells of depth E.
The distribution of the wells depth $\rho (E)$ is fixed a priori. P(E,t) is
the probability for a particle to be in a trap of depth E at time t. The
evolution of P(E,t) is simply governed by thermally activated hopping and
writes: 
\begin{equation}
\frac{\partial P(E,t)}{\partial t}=-P(E,t)e^{-E/T}+\Gamma (t)\rho (E)
\label{equevol}
\end{equation}

Where T is the thermal energy and $\Gamma (t)=\int_{0}^{\infty }P(E^{\prime
},t)e^{-E^{\prime }/T}dE^{\prime }$ is the average hopping rate. The time
unit is set to 1. Following \cite{monthus}, we take $\rho (E)=\exp (-E/T_{g})
$. For $T>T_{g}$, P(E,t) has a stationary limit $P(E,t)\propto \exp
[(1/T-1/T_{g})E]$. For $T<T_{g},$ P(E,t) has no stationary limit and keeps
evolving with time with a dynamics scaling as $\frac{t}{t_{w}}$. We solved
numerically equation \ref{equevol} for a quench from $T=\infty $ to $T=\frac{%
1}{2}T_{g}$. The energy distribution is presented in dotted line on fig 3b.
As expected it shifts progressively with time towards deeper and deeper
energy wells. The relaxation time of the system become thus longer and
longer. In order to mimic the strain sequence of fig 2, we now solve the
model for the following temperature history: The system is quenched from
infinite temperature to $T=\frac{1}{2}T_{g}$. After a delay of 100 time
units (tu) the temperature T is raised to $T=\frac{1}{2}T_{g}+\Delta T$ with 
$\Delta T=\frac{1}{3}T_{g}$, during 300 tu then quenched back to $\frac{1}{2}%
T_{g}$. t$_{w}$ is referenced after the second quench.The solution is
plotted in continuous line.

\begin{figure}[tbp]
\begin{center}
\includegraphics[width=8cm,clip]{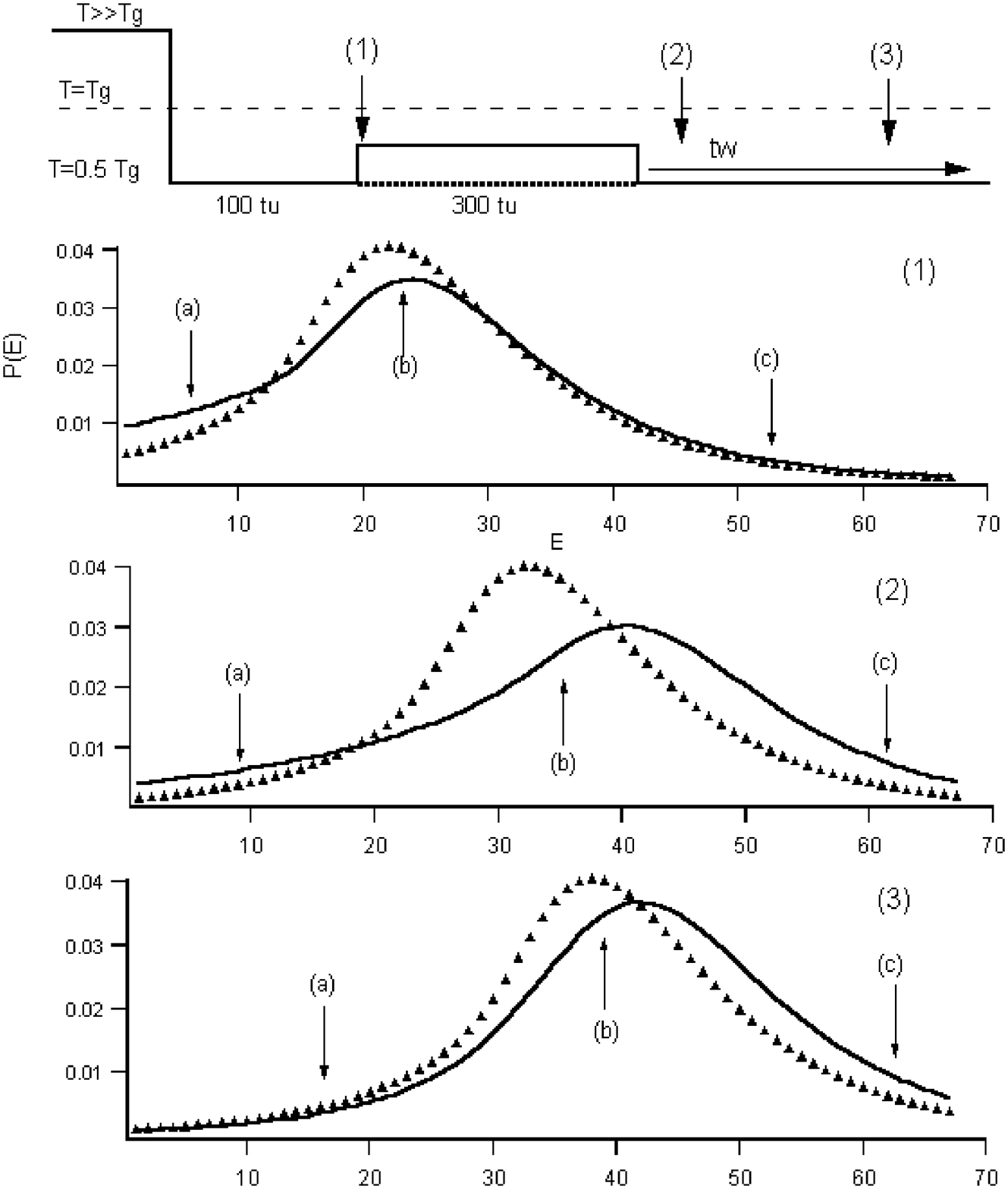}
\end{center}
\caption{a) Temperature history b) P(E,t$_{w}$) vs E for various t$_{w}.$
The ($\blacktriangle$) curve corresponds to the reference case; the full
line corresponds to the case with a step ($\Delta $T=$\frac{1}{3}$ T$_{g}$).
Notice that 1 tu after the sample is reheated (1) the small energies are
overpopulated, the intermediate ones are depleted and the large ones remain
unchanged. 1 tu after the second quench (2) both small and large energies
are overpopulated. 3000 tu after the second quench, small energies and
intermediate energies are depleted whereas large energies stay
overpopulated. }
\label{fig:PE}
\end{figure}

Fig 3b shows that shortly after the system is heated back (1) the small
energies -arrow (a) on fig 3- are overpopulated; intermediate energies (b)
are depleted and high energies (c) remain unperturbed compared to the
reference case where no temperature step is applied. The system ages then in
a higher temperature state. When the second quench happens (2) both low and
high energies are overpopulated whereas intermediate ones are depleted.
After a while (3) low energies recovered their reference population whereas
high energies stay overpopulated compare to the case without temperature
step. The system has consequently a longer average relaxation time.
Actually, we do not measure P(E,t) directly but we probe it experimentally
via $g_{1}(t+t_{w},t).$ The correlation function $g_{1}$ is a monotonically
increasing function of the probability that a particle has not changed trap
between $t_{w}$ and $t+t_{w}$. Within the frame of this model this
probability can be written: 
\[
C(t_{w}+t,t_{w})=\int_{0}^{\infty }P(E,t_{w})\exp [-t.e^{-E/T}]dE
\]
\begin{figure}[th]
\begin{center}
\includegraphics[width=8cm,clip]{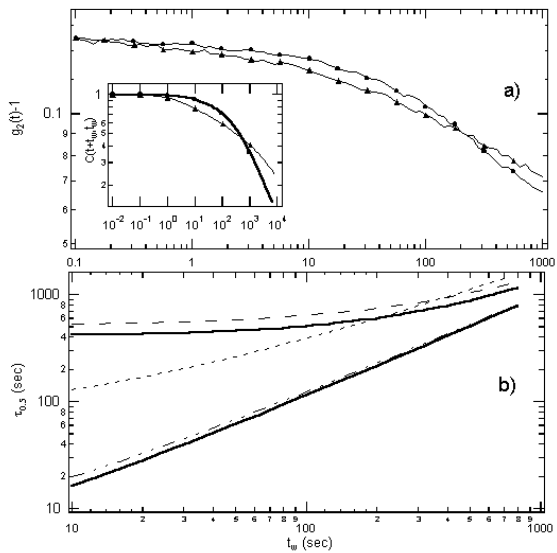}
\end{center}
\caption{a) $g_{2}(t_{w}+t,t_{w})-1$ for t$_{w}=1s$for $\Delta T=0$ ($%
\blacklozenge $) and for $\Delta T=\frac{1}{3}T_{g}(\triangle )$.The insert
shows the similar curves for $C(t_{w}+t,t_{w})$ calculated at t$_{w}=0.1$
tu. Notice that for $\Delta T=\frac{1}{3}T_{g}$ the curves first decrease
more rapidly ( small energies overpopulated ) then lies over the reference
one (large energies overpopulated). The agreement is qualitatively
excellent. b) $\tau _{1/2}$ calculated from $C(t_{w}+t,t_{w})$ vs t$_{w}$
for various $\Delta T=$0\% (bold line ), $\frac{1}{10}$ T$_{g}$ (----), $%
\frac{1}{3}$ T$_{g}$ (...), $\frac{3}{2}$ T$_{g}$ (-..-) and the reference
curve (bold line ). Notice the presence of overaging in the long time
regime. Qualitative agreement with fig 2b is satisfactory. }
\label{fig:Comparaison}
\end{figure}
Fig4(a) shows the change in the shape of the experimental correlation
function 1s after the shear cessation. It compares the case $\gamma =0$ (ref
case) with $\gamma =5.9\%$. Notice when a shear has been applied the decay
is faster at short times and becomes more slowly at long times. The
reference curve and the perturbed one can then cross each other. The insert
shows $C(t_{w}+t,t_{w})$ calculated at 0.1 tu after the temperature step. It
compares the case $\Delta T=0$ (ref case) with $\Delta T=\frac{1}{3}T_{g}$.
The modification in P(E,$t_{w})$ due to the step temperature is reflected in
the change of the correlation shape: the decrease is quicker at short times
(overpopulated low energies) and slower at long times (overpopulated high
energies). We observe an excellent qualitative agreement between the two set
of curves. This agreement is reinforced by fig 4(b) that shows the
calculated $\tau _{1/2}$ for different $\Delta $T. The calculated $\tau
_{1/2}$ are defined so that $C(t_{w}+\tau _{1/2},t_{w})=0.5.$ Fig 4b is
qualitatively similar to fig 2c. We point out that this phenomenon of
overaging in the trap model is robust to parameters changes. The values
presented here were chosen to obtain good looking curves but are not
critical at all. Notice that recent simulations \cite{BerthierBouchaud} on
Edward Anderson's model for spin glasses, show qualitatively the same
results for the correlation function with a positive temperature step.
Finally, we emphasize the fact that the change in the correlation function
shape reflects the change in the relaxation time distribution. We measure
how this distribution is transiently modified by shear. We expect that it
will provide an accurate selectivity on the models coupling mechanics and
thermal aging, but a more precise analysis with the existent models is
beyond the scope of this letter. We also point out that the qualitative
agreement with temperature step in the trap model reinforce the idea that
shear and temperature may play a similar role in certain glassy systems.We
thank J.P.Bouchaud, E.Bertin, C.Caroli, J.Kurchan, L.Berthier for fruitful
discussions. The authors thank M.Dorget for providing us with the latex beads.

\bibliographystyle{unsrt}

\end{document}